\documentclass[twocolumn,showpacs,preprintnumbers,amsmath,amssymb,showkeys]{revtex4}

\usepackage{array}
\usepackage{booktabs}
\usepackage{tabu}
\usepackage{dcolumn}
\usepackage{amsmath}
\usepackage{amsfonts}
\usepackage{amssymb}
\usepackage{graphicx}
\usepackage{subfigure}
\usepackage{graphicx}
\usepackage{dcolumn}
\usepackage{bm}

\def\be{\begin{equation}}
\def\ee{\end{equation}}
\def\bea{\begin{eqnarray}}
\def\eea{\end{eqnarray}}
\def\f{\frac}
\def\n{\nonumber}
\def\l{\label}
\def\p{\phi}
\def\o{\over}
\def\R{\rho}
\def\pa{\partial}
\def\om{\omega}
\def\na{\nabla}
\def\P{\Phi}

\begin{document}

\title{Hamilton-Jacobi formalism to warm inflationary scenario}

\author{K. Sayar}
\author{A. Mohammadi}
\email{abolhassanm@gmail.com}
\author{L. Akhtari}
\author{Kh. Saaidi}

\affiliation{
Department of Physics, Faculty of Science, University of Kurdistan, Sanandaj, Iran.\\}
\date{\today}

\def\be{\begin{equation}}
\def\ee{\end{equation}}
\def\bea{\begin{eqnarray}}
\def\eea{\end{eqnarray}}
\def\f{\frac}
\def\n{\nonumber}
\def\l{\label}
\def\p{\phi}
\def\o{\over}
\def\R{\rho}
\def\pa{\partial}
\def\om{\omega}
\def\na{\nabla}
\def\P{\Phi}

\begin{abstract}
Hamilton-Jacobi formalism as a powerful method is being utilized to reconsider warm inflationary scenario, where the scalar field as the main component driving inflation interacts with other fields. Separating the context to strong and weak dissipative regimes, the goal is followed for two popular functions of $\Gamma$. Applying slow-rolling approximation, the required perturbations parameters are extracted and by comparison to the latest Planck data, the free parameters are restricted. Possibility of producing an acceptable inflation is studied where the result shows that for all cases the model could successfully suggests amplitude of scalar perturbation, scalar spectral index, its running, and the tensor-to-scalar ratio.
\end{abstract}
\maketitle

\section{Introduction}
The most successful candidate describing the very early universe evolution is known as inflation. Besides solving the problems of Hot Big-Bang theory, inflation predicts cosmological perturbations in three types as scalar, vector, and tensor perturbations \cite{Linde,kolb,weinberg,mukhanov}. After proposing the inflationary scenario in 1981 by Alan Guth \cite{guth}, many different kinds of inflationary models have been introduced \cite{Linde2,brandenberger}.  \\
In a general view, the inflationary models could be classified in two categories as cold inflation and warm inflation. Cold inflationary scenarios are divided to two separate periods: inflation and reheating (or preheating). Since the scalar field is the dominant component of the universe and has no interaction with other fields, other components will be diluted due to accelerated expansion phase. After inflation ends, the universe stands in supercool phase and to restor hot big bang theory, reheating (or preheating) phase is necessary to heat up the universe and put it in radiation era \cite{kofman1,kofman2,greene,braden}. On the other hand, the mechanism of reheating is very crucial and it is very difficult to build up a consistent mechanism for reheating in every model of inflation. \\
An alternative to arrive at appropriate inflation is warm inflation \cite{berera,berera2,taylor,oliveira,oliveira2,hall,gil}. The the scalar field is still the dominant component however the main difference is that it has an interaction with the radiation field so that during the inflation period there is continuously a radiation production \cite{hall,oliveira2} through expansion, such that its energy density remains almost constant \cite{oliveira}. It is performed by dissipative coefficient that appears in the equation of motion of the scalar field and ensures an energy transfer from inflaton decay. Consequently the universe temperature does not drop intensely and the universe is able to smoothly cross to the radiation phase. Then there is no need for any reheating period \cite{hall,oliveira2,cid}. As a further argument, it should be noted that the radiation temperature with $T>H$ is a necessary condition for warm inflation to happen \cite{herrera,herrera2}. The point would be clear as one attends to the fact that thermal and quantum perturbation are respectively dependent on $T$ and $H$ \cite{berera,hall,berera3,berera4}. As a result, as long as $T>H$ the quantum fluctuation is defeated by thermal fluctuation \cite{herrera2}. Then, it is determined by warm inflation that how thermal fluctuation could play the role of dominant initial fluctuation and be a seed for Large-Scale Structure of the Universe. Therefore, it could be concluded that the density fluctuation arises from thermal fluctuation; where they have their origin in hot radiation and the scalar field is affected by them through the friction term existed in the scalar field equation of motion \cite{berera2,berera3,berera5,hall2}. \\
There are literatures on warm inflationary scenario studying different aspects of the scenario \cite{campo,campo2,campo3,cid,Visinelli}. In the presented work, the main aim is to consider warm inflationary scenario utilizing a different formalism named Hamilton-Jacobi formalism. Instead of the potential, in the formalism Hubble parameter is given as a function of the scalar field. This formalism in cold inflationary studies has got its place \cite{salopek,liddle,kinney,guo,aghamohammadi,saaidi,sheikhahmadi}, however up to our knowledge there is no work considering warm inflation using this formalism. Then it sounds interesting to investigate warm inflation using this powerful formalism. \\
The paper has been organised as follows: In Sec.II, the general dynamical equations of the model will be introduced, and the slow-rolling parameters are obtained in agreement with the formalism. The situation could be consider for two separate regimes so that Sec.III belongs to the strong dissipative regime, where $\Gamma \gg 3H$, and weak dissipative regime will be investigated in Sec.IV. Utilizing the perturbation relations, the free parameters of the model will be restricted such that the model predictions be in agreement with data. The obtained potential for each case is depicted that shows almost the same behavior as the potential for chaotic inflation where the the scalar field slowly rolls down to the bottom of its potential. Additionally, for a more comparison, the energy densities of the scalar field and radiation are illustrated and the behavior is explained properly. The result of the work is gathered in the last section. \\

\section{Generality}
\label{Sec2}

Consider a spatially flat space-time universe describing by well-known FLRW metric and filled with a perfect fluid of energy density $\rho_r$ and a self-interacting scalar field which drives the inflation and possesses the energy density $\rho_\phi = \dot\phi^2/2 + V(\phi)$. The evolution of the Universe could be stated by Friedmann equation as follows
\begin{equation}\label{Friedmann}
H^2 = {1 \over 3M_p^2} (\rho_\phi + \rho_r)
\end{equation}
The scalar field is in interacting with the other component of the Universe, so that its evolution is expressed as
\begin{equation}
\ddot\phi + 3H(1+Q)\dot\phi + V'= 0
\end{equation}
where the parameter $Q$ is defined by $Q\equiv \Gamma /3H$ which is the ratio of radiation production rate to expansion rate, and $\Gamma$ is the dissipative coefficient. On the other hand, the radiation field is govern by the transfer relation
\begin{equation}
\dot{\rho}_r + 4H\rho_r=\Gamma \dot{\phi}^2
\end{equation}
In addition, it is assumed that during the inflation, the radiation production is quasi-stable i.e $\dot\rho_r \ll 4H\rho_r, \Gamma\dot\phi^2$. Then, the radiation energy density could be approximated by
\begin{equation}\label{rhor}
\rho_{r}=\alpha T^4 = {\Gamma \over 4H}\; \dot\phi^2
\end{equation}
where $T$ denotes the radiation temperature, $\alpha \equiv \pi^2 g_\ast / 30$ is the Stefen-Boltzman constant, and $g_\ast$ is the number of degree of freedom for radiation field that in the standard cosmology is taken of order $100$.  \\
To estimate the time derivative of the scalar field, we work with the second Friedmann equation given by
\begin{equation}\label{Friedmann2}
\dot{H} = -{1 \over 2M_p^2} \; \big( 1+Q  \big) \dot\phi^2.
\end{equation}
In the Hamiltonian-Jacobi formalism, the Hubble parameter is given as a function of the scalar field. Therefore, the time derivative of the Hubble parameter reexpressed by $\dot{H}=\dot\phi H'$, and consequently, $\dot\phi$ is derived as
\begin{equation}\label{phidot}
\dot\phi = -{2M_p^2 \over (1+Q)}\; H'(\phi)
\end{equation}
and the potential of the scalar field is predicted from Eq.(\ref{Friedmann})
\begin{equation}\label{pot}
V(\phi) = 3M_p^2H^2(\phi) - {2M_p^4 \over (1+Q)^2}H'^2(\phi)
\end{equation}
The last two equations are known as the Hamiltonian-Jacobi equations. \\
Necessity of having an accelerated expansion during the inflation leads one introduce the dimensionless parameter $\epsilon \equiv -{\dot{H} \over H^2}$, known as the first slow-rolling parameter that could be rewritten as
\begin{equation}\label{epsilon}
\epsilon(\phi) = -{\dot{H} \over H^2} = {2M_p^2 \over (1+Q)}\; {H'^2(\phi) \over H^2(\phi)}
\end{equation}
Smallness of $\epsilon(\phi)$ indicates that during inflation the kinetic part of the scalar field energy density is much smaller that the potential part. Furthermore, by attention to the Eqs.(\ref{rhor}), (\ref{phidot}) and (\ref{epsilon}) one could reexpress the radiation energy density and find the following expression between the scalar field energy density and radiation energy density
\begin{equation}\label{renergy}
\rho_r = {Q \over 2(1+Q)} \; \epsilon \rho_\phi,
\end{equation}
so that at the end of inflation where $\epsilon=1$, there is $\rho_r = {Q \over 2(1+Q)} \; \rho_\phi$. \\
The second slow-rolling parameter defined as $\epsilon_2 \equiv -\ddot{H}/H\dot{H}$, that results in the following expression 
\begin{equation*}
\epsilon_2 = \eta(\phi) - {Q \over 1+ Q} \Big( \beta(\phi) - \epsilon(\phi) \Big)
\end{equation*}
with the following definitions
\begin{equation}\label{eta}
\eta(\phi) \equiv {4M_p^2 \over (1+Q)}\; {H''(\phi) \over H(\phi)}, \qquad
\beta \equiv {2 M_p^2 \over (1+Q)}\; {\Gamma' H' \over \Gamma H}.
\end{equation}
In addition to the above slow-rolling parameters, there are other parameters which will appear in the calculation of the next sections
\begin{equation*}
\sigma \equiv {2 M_p^2 \over (1+Q)}\; {H''' \over  H'}, \qquad
\delta \equiv {4 M_p^4 \over (1+Q)^2}\; {\Gamma'' H'^2 \over \Gamma H^2}.
\end{equation*}
Amount of inflation is very important for solving the problems of the big-bang model. This could be measured by number of e-fold given by
\begin{equation}\label{efold}
N(\phi) ={1 \over 2M_p^2} \int_{\phi_e}^{\phi} (1+Q){H(\phi) \over H'(\phi)} d\phi
\end{equation}
where the subscribe "e" stands for the end of inflation. \\
To test a proposed model, it should be compared with the observational data. The most important data that could be used in inflationary scenario are those comes from perturbation theory such as the amplitude of scalar perturbation, the scalar spectral index and so on. Following \cite{taylor}, the amplitude of scalar perturbation is derived as
\begin{equation}\label{scalarperturbation}
P_s = {25 \over 4} \; {H^2 \over \dot\phi^2}\; \delta\phi^2
\end{equation}
and the scalar spectral index and the running of scalar spectral index come directly from $\mathcal{P}_s$
\begin{equation}\label{ns}
n_s - 1 = {d\ln\big(\mathcal{P}_s \big) \over d\ln(k)}, \qquad \alpha_s = {dn_s \over d\ln(k)}.
\end{equation}
Another prediction of inflationary scenario is the tensor perturbation which is detected indirectly, and measured by the tensor-to-scalar ratio parameter $r=\mathcal{P}_t / \mathcal{P}_s$.  \\
Analysing the model needs first to specify the Hubble parameter as a function of the scalar field. It is assumed that the Hubble parameter is a power-law function as $H(\phi)=H_0\phi^n$. The main reason for this choice comes from Planck data, where it is realized that the power-law potentials could be in good agreement with observational data \cite{planck}. On the other side, according to first Hamilton-Jacobi equation, it is clear that a power-law function of the Hubble parameter leads to a power-law function for the potential. Consequently, it seems that the choice for the Hubble parameter could be a reasonable selection. Before taking the next step, we separate the work into strong and weak dissipative regime, then investigate the situation for different types of dissipative coefficient. Generally speaking, this quantity could be a function of scalar field, or in general view a function of scalar field and temperature. In order to arrive at better insight and understanding about warm inflationary scenario, both of the possibilities have been taken. In the first case, the parameter $\Gamma$ is taken as a power-law function of scalar field, and for the the second case it is assumed as the well-known expression $T^m / \phi^{m-1}$. The main motivation could be explained as following lines:\\
In \cite{first,first01,hall2}, the authors took an interaction as $\lambda\phi^2\chi^2 / 2$ and $g\chi\bar{\psi}\psi$, in which the inflationary period presents a two-stage decay chain $\phi  \chi   \psi$. They proved that the dissipation coefficient $\Gamma$ becomes $\lambda^3 g^2 \phi /256 \pi^2$ that could be considered as a power-law function $\propto \phi^m$ with $m=1$. In the presented work, it is taken more generally, and $m$ is left undetermined. Then, we use Planck data to realized which $m$ could be the best choice. \\
The other choice for the dissipation coefficient, which is a more general form, is the one that $\Gamma$ is a function of both the scalar field and temperature, namely $\Gamma = C_\phi T^m / \phi^{m-1}$ \cite{second,second2,second3,second4,second5,second6}. This choice is more interesting than the first one. The case with $m=1$ corresponds to $\Gamma \propto T$ describing the high temperature supersymmetry case; $m=0$ the dissipation coefficient is only depends on the scalar field, $\Gamma \propto \phi$ that represents an exponentially decaying propagator in the supersymmetry case; and also for $m=-1$ there is $\Gamma \propto \phi^2 /T$ that corresponds to the non-supersymmetry case. Moreover, the case with $m=3$, namely $\Gamma \propto T^3 / \phi^2$ was studied in \cite{del-Campo}. The generality of this case was a motivation for us to pick out it and study the corresponding result.\\

\section{Consistency of the model in Strong Dissipative Regime}
In this regime the parameter $Q$ is bigger than unity, $Q \gg 1$ in which the coefficient $(1+Q)$ that appears in the main equations could be approximated as $(1+Q) \simeq Q$. On the other hand, the fluctuations in the inflation is produced by the thermal fluctuations instead of the quantum fluctuations, which results in $\delta\phi^2=k_FT/2\pi^2$ where $k_F=\sqrt{\Gamma H}$ \cite{berera2,taylor,herrera,herrera2}. Substituting the term in Eq.(\ref{scalarperturbation}) comes to the amplitude of scalar perturbation for strong dissipative regime as
\begin{equation}\label{strongSP}
\mathcal{P}_s= {25 \over 32\pi^2 M_p^3} \left[ \Gamma^3 H \over 9 \alpha^{1/3} H'^2 \right]^{3 \over 4}.
\end{equation}
Then, the scalar spectral index and its running could be read as

\begin{equation}\label{strongns}
n_s = 1 - {3 \over 4}\; (\epsilon - \eta + 3 \beta),
\end{equation}
\begin{equation}\label{strongrunning}
\alpha_s = {3 \over 4} \left( \epsilon\eta - \epsilon^2 - 2\epsilon\sigma + 3\delta -6\beta^2 + {5 \over 2}\beta\eta - \epsilon\beta \right).
\end{equation}
Another important perturbation parameter which implies on tensor perturbation is the tensor-to-scalar ratio that in our case is given by
\begin{equation}\label{strongr}
r = {\mathcal{P}_t \over \mathcal{P}_s} = {64 \over 25} \left( 3\alpha^{1/3} \over 2 \right)^{3 \over 4} \; {H^2 \over \sqrt{M_p} \; \Gamma^{3 \over 2}} \; \epsilon^{3/4}.
\end{equation}
Going forward and reaching more detail requires to apply a specific dissipation coefficient. At the following subsections, the situation will be studied for two typical examples of the dissipation coefficient, and the parameters of the model will 
be restricted by comparison to the observational data.  \\

\subsection{$\Gamma := \Gamma(\phi)$}
\begin{figure}
  \centering
  \includegraphics[width=7cm]{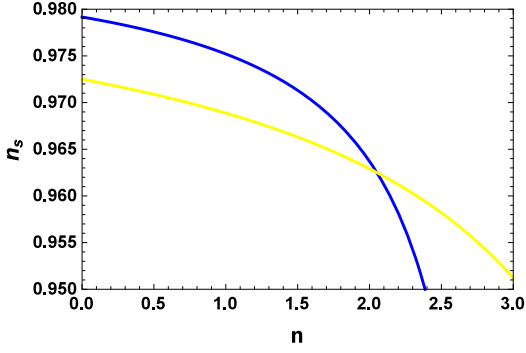}
  \caption{Scalar spectral index versus $n$ for $N=60$ and different values of $m$ as $m=1$ (blue color line) and $m=3$ (yellow color line).}\label{sns01}
\end{figure}

\begin{figure}
  \centering
  \includegraphics[width=7cm]{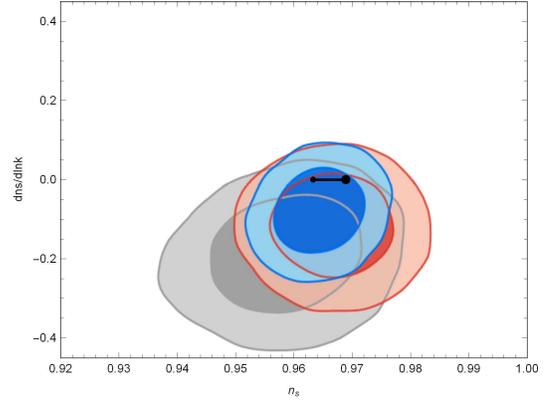}
  \caption{The running of scalar spectral index is shown versus $n_s$ for $n=1.5$ and $m=3$, and for $N=55-65$ in which the small point belongs to $N=55$ and the large point belongs to $N=65$.}\label{sasns01}
\end{figure}

\begin{figure}
  \centering
  \includegraphics[width=7cm]{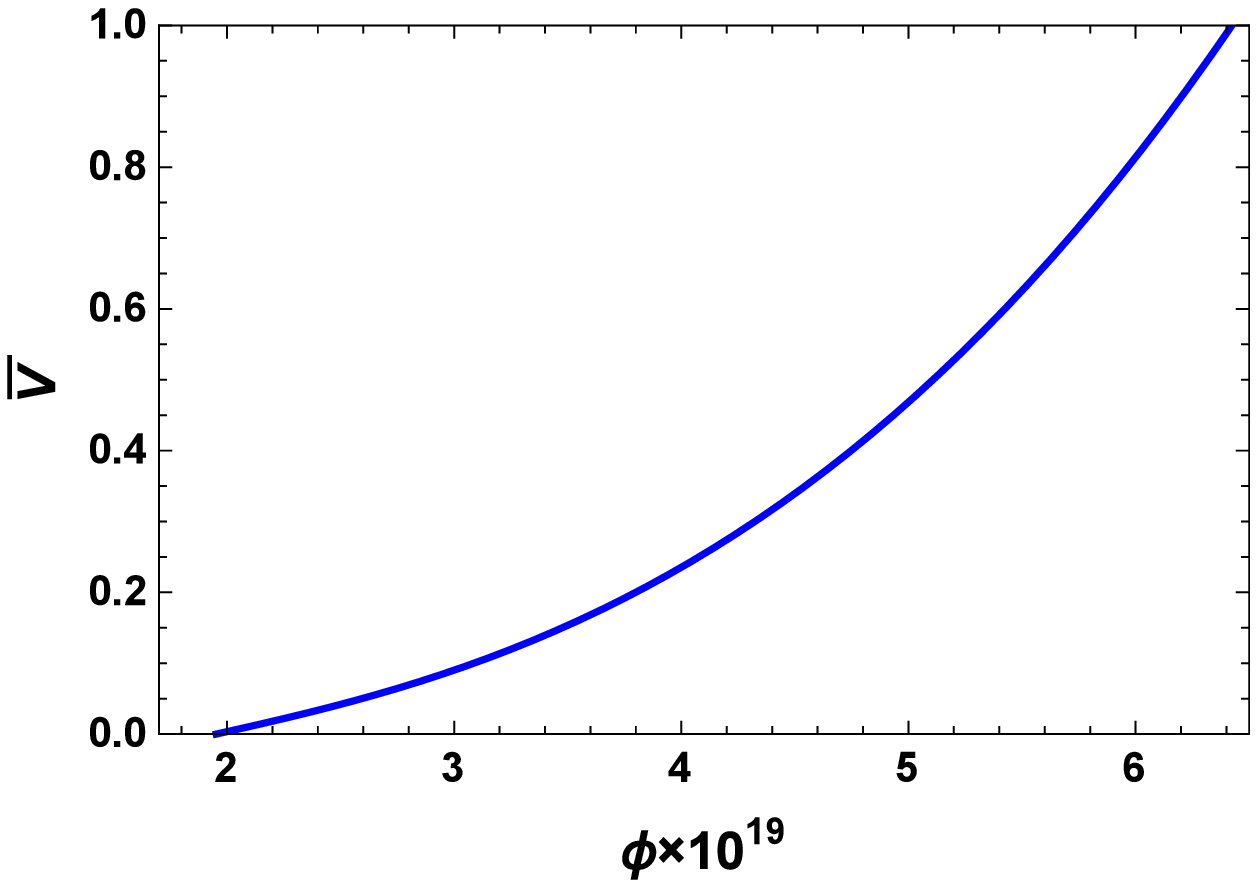}
  \caption{The potential versus the scalar field is plotted during the inflationary era, where the constant parameters are taken as: $n=1.5$, $m=3$, $N=60$, and $H_0=3\times10^{-18}$. The parameter $\bar{V}$ is defined by $\bar{V}={V(\phi) \over V_0}$, in which $V_0=4.23\times 10^{61}$ is the initial value of the potential.}\label{spot01}
\end{figure}

\begin{figure}
  \centering
  \includegraphics[width=7cm]{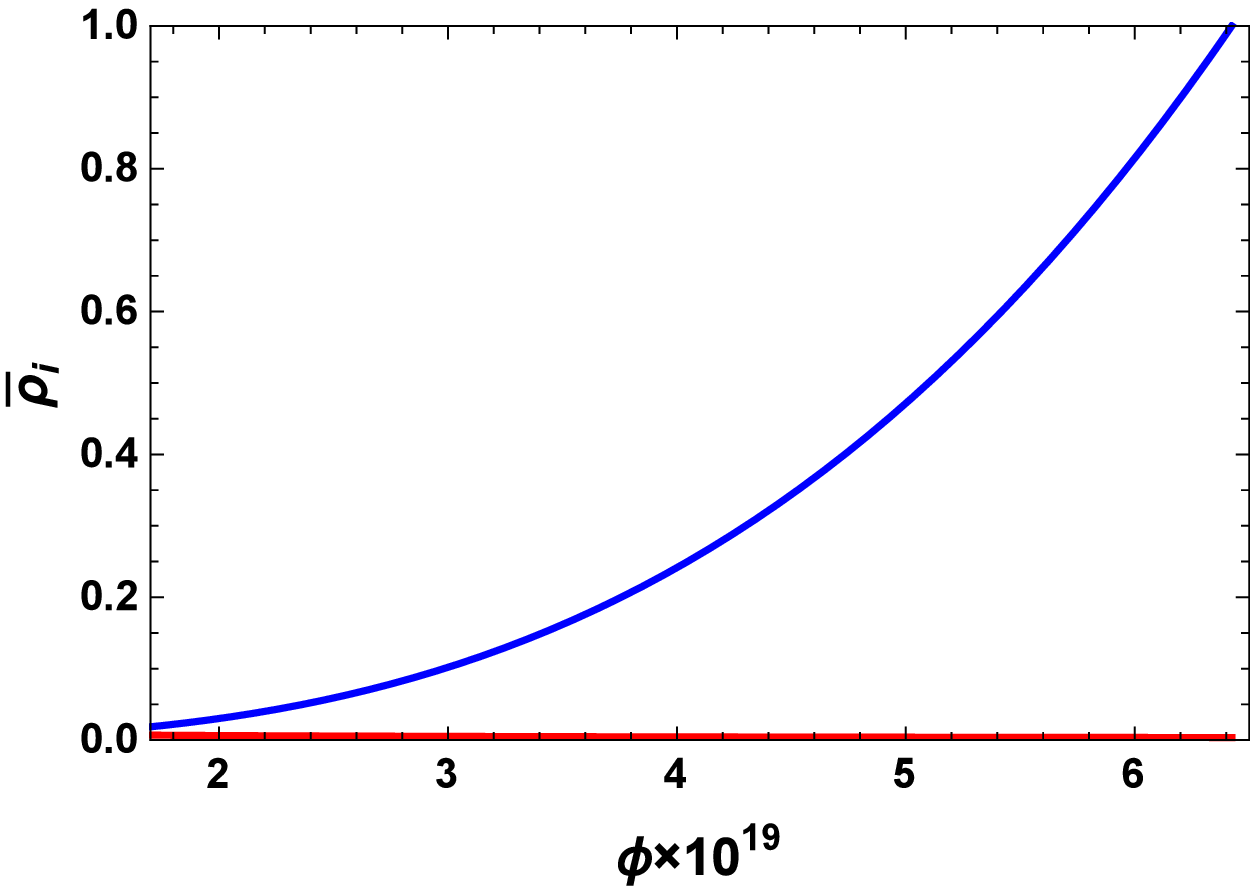}
  \caption{Energy densities of the scalar field (blue line) and radiation (red line) is depicted in term of the scalar field during the inflation times, where the constant parameters are taken as: $n=1.5$, $m=3$, $N=60$, and $H_0=2\times10^{-18}$. The parameter $\bar{\rho}_i$ is defined by $\bar{\rho}_i={\rho_i(\phi) \over \rho_0}$, in which $\rho_0$ is the initial value of the scalar field energy density.}\label{senergy01}
\end{figure}

As a first case, the dissipative coefficient is taken as a power-law function of the scalar field, $\Gamma=\Gamma_0 \phi^m$, where $\Gamma_0$ and $m$ are constant. Inflation is defined as an accelerated expansion phase of the very early Universe evolution. Naturally, 
the inflation period ends as the acceleration gets $\ddot{a}=0$ or equivalently as $\epsilon$ arrives at unity, $\epsilon=1$. From Eq.(\ref{epsilon}), the scalar field at the end of inflation could be read. Then, utilizing the number of e-fold (\ref{efold}), the scalar field at the time when the perturbations exit the horizon gets the following expression
\begin{equation}\label{phii01}
\phi_\ast^{2+m-n} = \phi_e^{2+m-n} \left( 1 + {2+m-n \over n} \; N \right),
\end{equation}
in which
\begin{equation*}
\phi_e^{2+m-n} = {6n^2M_p^2 H_0 \over \Gamma_0}.
\end{equation*}
Based on observational data the scalar spectral index should be about $n_s=0.9677 \pm 0.0060$ ($68\%$ CL, \textit{Planck} $ \rm{TT+lowP+lensing}$), and for its running there is $\alpha_s = -0.0033 \pm 0.0074$ ($68\%$ CL, \textit{Planck} $ \rm{TT+lowP+lensing}$) \cite{planck}. The model anticipates the slow-rolling parameters at horizon crossing as
\begin{eqnarray*}
 \epsilon^\ast &=& {n \over n+(2+m-n)N}, \\
 \eta^\ast &=& {2(n-1) \over n}\epsilon^\ast , \\
 \beta^\ast &=& {m \over n}\epsilon^\ast , \\
 \sigma^\ast &=& {(n-1)(n-2) \over n^2}\epsilon^\ast, \\
 \delta^\ast&=& {m(m-1) \over n^2}\epsilon^{\ast 2}.
\end{eqnarray*}
Therefore, both $n_s$ and $\alpha_s$ at horizon crossing could be derived as  a function of $n$ and $m$. These free parameters should be picked out in a way to put $n_s$ in acceptable range. To arrive at this purpose, $n_s$ is depicted versus $n$ for different values of $m$ in Fig.\ref{sns01}, so that the perfect values for the parameters could be chosen with high confidence. It is realized that by growing $m$ the parameter $n_s$ drops slower, has a smaller slope, and starts from smaller amount.  \\
Obviously, $(n,m)=(1.5,3)$ sounds to be a reasonable choice, where by applying it, the scalar spectral index and the running at horizon 
crossing are acquired as $n_s = 0.9663$ and $\alpha_s=-8.59\times 10^{-4}$ so that they are both in perfect agreement
with observational data. To check the result more carefully, $\alpha_s$ versus $n_s$ has been plotted and for having a thoughtful comparison it has been illustrated on the Planck data as Fig.\ref{sasns01}. The black line is the model prediction for $N=55-65$ that 
stands in the $68\%$ CL area completely displaying the consistency of the model.
On the other hand, the latest observational data implies that the amplitude of scalar perturbation at horizon crossing is about $\ln\Big( 10^{10}\mathcal{P}_s^\ast \Big) = 3.062 \pm 0.029$, and the tensor-to-scalar ration $r$ has an upper bound $r^\ast <0.11$ at $68\%$ CL \cite{planck}. These data could be applied to confined the free parameters of the model as
\begin{eqnarray*}
  H_0 &=& \left[ {\pi^2 M_p^2 r^\ast \mathcal{P}_s^\ast \over 2 \left( 6n^2 M_p^2 \big[ 1+ {2+m-n \over n}N \big] \right)^{2n \over 2+m-n}} \right]^{2-n \over 4} \; C^{n \over 2(2+m-n)}, \\
  \Gamma_0 &=& \left( {C \over H_0^m} \right)^{1 \over 2-n},
\end{eqnarray*}
where
\begin{equation*}
  C = \left[{ {288 \over 25}\pi^2 n^{3 \over 2}M_p^3 \sqrt{\sqrt{\alpha} \over 3} \; \mathcal{P}_s^\ast   \over   \left( 6n^2 M_p^2\big[ 1 + {2+m-n \over n}\; N \big] \right)^{3(3m+2-n) \over 4(m-n+2)} }\right]^{2(m-n+2) \over 3}.
\end{equation*}
Consequently, selecting an appropriate values for the above parameter ensures that the tensor-to-scalar ratio $r$ also stands in agreement with data. \\
The potential behavior could be realized according to Eq.(\ref{pot}) in which for consistent values of the model free parameter it is exhibited in Fig.\ref{spot01}. The potential sounds to have a similar behavior as chaotic potential, in which the scalar field rolls down slowly from top of the potential and moves toward the minimum of it. Further, it should be noted that during the inflation times the scalar field takes values smaller than the Planck mass and the potential energy is below the Planck energy too.
Also, based on Eq.(\ref{Friedmann}) and (\ref{rhor}), one could easily compare the energy densities of the scalar field and radiation which has been plotted in Fig.\ref{senergy01}. It is observed that the scalar field energy density at onset of inflation is the dominant component. By passing time it decreases however radiation energy density remains almost constant, and at the end of inflation one could find out that both component possess comparable values.

\subsection{$\Gamma := \Gamma(\phi,T)$}
\begin{figure}
  \centering
  \includegraphics[width=7cm]{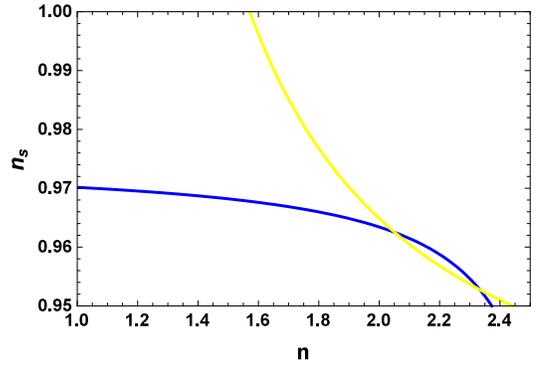}
  \caption{Scalar spectral index versus $n$ for $N=60$ and different values of $m$ as $m=-1$ (blue color line) and $m=4$ (yellow color line).}\label{sns02}
\end{figure}

\begin{figure}
  \centering
  \includegraphics[width=7cm]{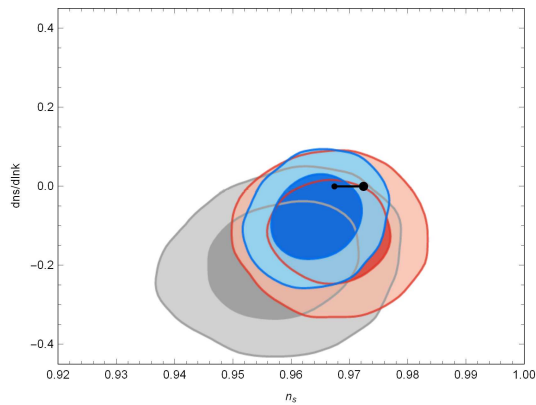}
  \caption{The running of scalar spectral index is shown versus $n_s$ for $n=1$ and $m=-1$, and for $N=55-65$ in which the small point belongs to $N=55$ and the large point belongs to $N=65$.}\label{sasns02}
\end{figure}

\begin{figure}
  \centering
  \includegraphics[width=7cm]{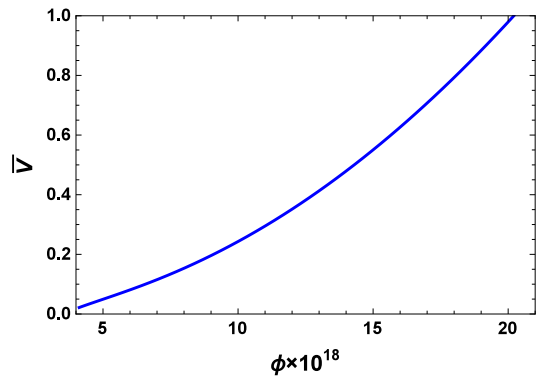}
  \caption{The potential versus the scalar field is plotted during the inflationary era, where the constant parameters are taken as: $n=1$, $m=-1$, $N=60$, and $H_0=1\times 10^{-8}$. The parameter $\bar{V}$ is defined by $\bar{V}={V(\phi) \over V_0}$, in which $V_0=7.26\times 10^{59}$ is the initial value of the potential.}\label{spot02}
\end{figure}

\begin{figure}
  \centering
  \includegraphics[width=7cm]{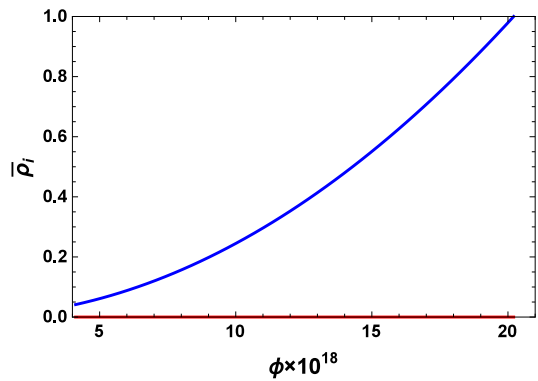}
  \caption{Energy densities of the scalar field (blue line) and radiation (red line) is depicted in term of the scalar field during the inflation times, where the constant parameters are taken as: $n=1$, $m=-1$, $N=60$, and $H_0=1\times 10^{-8}$. The parameter $\bar{\rho}_i$ is defined by $\bar{\rho}_i={\rho_i(\phi) \over \rho_0}$, in which $\rho_0$ is the initial value of the scalar field energy density.}\label{senergy02}
\end{figure}

The more general dissipative coefficient is taken for this case where it is assumed as a function of the scalar field and 
temperature, $\Gamma=\Gamma_0 {T^m \over \phi^{m-1}}$. From Eq.(\ref{rhor}), temperature could be evaluated as a function of the scalar field
\begin{equation}\label{temp2}
  T(\phi) = \left( {9M_p^4 n^2 H_0^3 \over \alpha \Gamma_0} \right)^{1 \over 4+m} \; \phi^{3n+m-3 \over 4+m}.
\end{equation}
Therefore, one arrives at a definition in term of the scalar field for the dissipative coefficient
\begin{equation}\label{gamma2}
  \Gamma(\phi) = \left( {9M_p^4 n^2 \over \alpha} \Gamma_0^{4/m} H_0^3 \right)^{m \over 4+m} \phi^{3nm-6m+4 \over 4+m}.
\end{equation}
Since the process is same as the previous case, we briefly explain the detail and try to get to the main result. The scalar field at the horizon exit time is read by
\begin{equation}\label{phii2}
  \phi_\ast^{2\nu \over 4+m} = \phi_e^{2\nu \over 4+m} \Big( 1 + {2\nu \over n(4+m)}\; N \Big),
\end{equation}
so that the final scalar field is given by
\begin{equation*}
  \phi_e^{2\nu \over 4+m}= {6n^2 M_p^2 H_0 \over \Big( {9M_p^4 n^2 \over \alpha} \Gamma_0^{4/m} H_0^3 \Big)^{m \over 4+m}}.
\end{equation*}
where $\nu=mn-2n-2m+6$. The slow-rolling parameters $\epsilon^\ast$, $\eta^\ast$, and $\sigma^\ast$ are derived same as before, and the only difference is presented in $\beta^\ast$ and $\delta^\ast$, 
\begin{equation*}
\beta^\ast={3mn-6m+4 \over n(4+m)} \epsilon^\ast, \quad
\delta^\ast = {(3mn-7m)(3mn-6m+4) \over n^2(4+m)^2} \epsilon^{\ast 2}.
\end{equation*}
and by inserting them in Eqs.(\ref{strongns}) and (\ref{strongrunning}), the scalar spectral index and running are derived in term of $n$ and $m$.
Fig.\ref{sns02} portrays the scalar spectral index versus $n$ for different values of $m$, that could be used to select a proper values. By increasing $m$, $n_s$ drops faster with higher slope. However, when negative values for $m$ is applied it could be found out that $n_s$ start at smaller values and changes slower so that there is a bigger range of $n$ resulting in the right value of $n_s$.
$(n,m)=(1,-1)$ plays a proper choice that comes to $n_s=0.9701$ and $\alpha_s=-6.68\times 10^{-4}$ where they both are consistent with observational data. To get a better conclusion for the case, the running of scalar spectral index is depicted versus $n_s$ as in Fig.\ref{sasns02} where one can easily get more evidence about the model prediction. The result is acceptable however for $N=65$ it goes out of $68\%$ CL.
Then utilizing the observational data for the amplitude of the scalar field and the tensor-to-scalar ratio parameters comes to the following expression for the free parameters of the model
\begin{eqnarray*}
  H_0 &=& \left[{ \pi^2 M_p^2 r^\ast \mathcal{P}_s^\ast \; C^{4n \over \mu\nu }    \over
                 2 \Phi^{n(4+m) \over \nu} }\right]^{1 \over p}, \\
  \Gamma_0 &=& \left( {C \over H_0^{4+m}} \right)^{1 \over \mu}.
\end{eqnarray*}
in which the constant quantities are defined by
\begin{eqnarray*}
  C & \equiv & \left[ { {288 \over 25} \pi^2 n^{3/2} M_p^3 \sqrt{\sqrt{\alpha} \over 3} \mathcal{P}_s^\ast     \over
                 \Big( {9n^2 M_p^4 \over \alpha } \Big)^{9m \over 4(4+m)} \Phi^{3z \over 2\nu}  } \right]^{\nu(4+m) \over 3}, \\
  p & \equiv & 2 + {2n(2-m) \over \nu} + {4n(4+m) \over \mu\nu}, \\
  \mu & \equiv & 8+2m-4n-nm, \\
  z & \equiv & 2nm-4m-n+5, \\
  \Phi & \equiv & 6 n^2 M_p^2 \Big( { \alpha \over 9n^2 M_p^4 } \Big)^{m \over 4+m} \Big[ 1 + {2\nu \over n(4+m)}\; N \Big].
\end{eqnarray*}
The potential is required to be below the Planck energy for the choice. Therefore, the potential behavior is illustrated in Fig.\ref{spot02} versus the scalar field during the inflation period, which determines that the potential is below the Planck energy. The potential has a similar behavior as before where the scalar field slowly falls down to the minimum area. By comparison to the previous case, the scalar field again takes a values smaller than the Planck mass and also much smaller than previous case. 
In addition, the dominance of the scalar field on radiation could be found out by plotting and comparing both energy densities. Fig.\ref{senergy02} shows both energy densities in term of the scalar field during the inflationary times indicating that at the beginning of inflation the scalar field energy density is much bigger than the radiation energy density. Then, by passing time they come closer to each other.

\section{Consistency of the model in Weak Dissipative Regime}
For this case, the parameter $Q$ is much smaller than unity so that in the main relation we use the approximation $1+Q\simeq 1$. The 
scalar field perturbation is approximated about $\delta\phi \simeq HT$ \cite{herrera,herrera2}, so the amplitude of scalar perturbation as the most important equation in the work will be different from the previous case and one expects changes. Inserting $\delta\phi$ in Eq.(\ref{scalarperturbation}), the amplitude of scalar perturbation for the regime could be derived as
\begin{equation}
\mathcal{P}_s = {25H^3 \over 16M_p^3 H'^2}\; \left( {\Gamma H'^2 \over \alpha H} \right)^{1 \over 4}.
\end{equation}
Consequently, taking derivative of above relation, the scalar spectral index and its running could be acquired respectively as
\begin{equation}\label{weakns}
n_s=1 - \dfrac{1}{4}\Big( 11 \epsilon - 3 \eta + \beta \Big)
\end{equation}
\begin{equation}\label{weakas}
\alpha_s =  {1\over 2} \left( 7 \epsilon\eta - 11\epsilon^2 - 3\epsilon\sigma + {1 \over 4} \eta\beta - {1 \over 2} \epsilon\beta + {1 \over 2}\delta - {1 \over 2}\beta^2 \right)
\end{equation}
It is clear that the slow-rolling parameter $\epsilon$ and number of e-folds $N$ do not depend on $Q$. Then, the scalar field at the beginning and end of inflation has the same value for any choice of dissipative coefficient $\Gamma$, namely
\begin{equation}\label{phiiw}
\phi_\ast^2 = \phi_e^2 \Big( 1 + {2 \over n}N \Big),  \quad \qquad \phi_e^2= 2n^2 M_p^2.
\end{equation}
Gravitational waves as another prediction of inflation could be identified by measuring the the tensor-to-scalar ratio parameter $r$, that in our case is obtained as
\begin{equation}
r =  {\mathcal{P}_t \over \mathcal{P}_s} = {16 \over 25\pi^2 \sqrt{M_p}} \; \left( {2\alpha H^3 \over \Gamma} \right)^{1 \over 4} \; \epsilon^{3/4}.
\end{equation}
To get some insight, the situation will be studied in the following subsections for two choices of dissipative coefficient. \\

\subsection{$\Gamma := \Gamma(\phi)$}
\begin{figure}
  \centering
  \includegraphics[width=7cm]{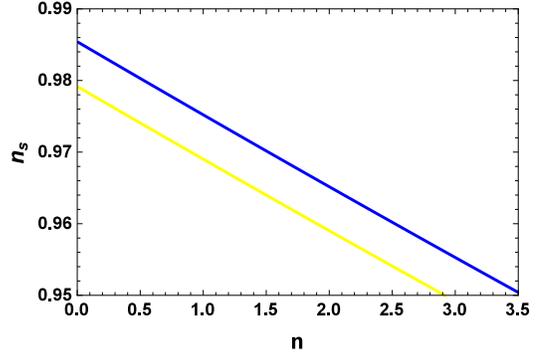}
  \caption{Scalar spectral index versus $n$ for $N=60$ and different values of $m$ as $m=1$ (blue color line) and $m=4$ (yellow color line).}\label{wns01}
\end{figure}

Consider the dissipative coefficient as a power-law function of the scalar field $\Gamma=\Gamma_0 \phi^m$. Utilizing it, and getting the initial scalar field from (\ref{efold}), the slow-rolling parameters at the horizon exit are obtained as
\begin{eqnarray*}
\epsilon^\ast &=& {n \over 2N+n},  \\
\eta^\ast &=& {2(n-1) \over n}\epsilon^\ast, \\
\sigma^\ast &=& {(n-1)(n-2) \over n^2}\epsilon^\ast, \\
\beta^\ast &=& {m \over n}\epsilon^\ast, \\
\delta^\ast &=& {m(m-1) \over n^2 } \epsilon^{\ast 2}.
\end{eqnarray*}
Then, substituting above parameters into Eqs.(\ref{weakns}) and (\ref{weakas}) the scalar spectral index and the running are derived in 
term of $n$ and $m$. Plotting the parameter $n_s$ as in Fig.\ref{wns01}, it is realized that for proper choice of $n$ and $m$, $n_s$ could stand in acceptable range predicted by Planck data. It sounds that for different values of $m$, $n_s$ diagram almost has the same slope. Also by increasing $m$, $n_s$ gets smaller values. \\
Taking $n=2$ and $m=1$, the scalar spectral index and its running are estimated respectively as $n_s=0.9651$ and $\alpha_s=-5.71\times 10^{-4}$ which clearly sounds that is in perfect agreement with observational data. It is still required more attention and in this regard the parameter $\alpha_s$ is illustrated in term of $n_s$ in Fig.\ref{wasns01} where one could find that the whole line stands in $68\%$ CL which is a statement of the model agreement with data.
\begin{figure}
  \centering
  \includegraphics[width=7cm]{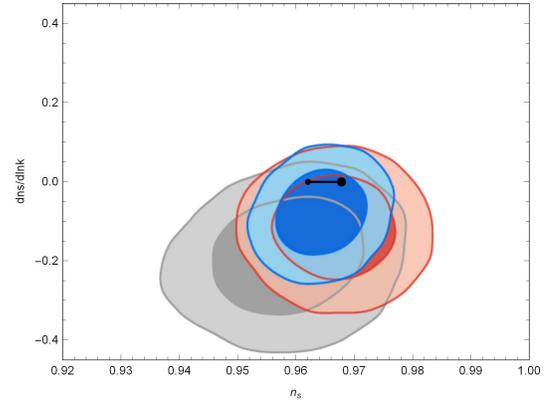}
  \caption{The running of scalar spectral index is shown versus $n_s$ for $n=2$ and $m=1$, and for $N=55-65$ in which the small point belongs to $N=55$ and the large point belongs to $N=65$.}\label{wasns01}
\end{figure}

\begin{figure}
  \centering
  \includegraphics[width=7cm]{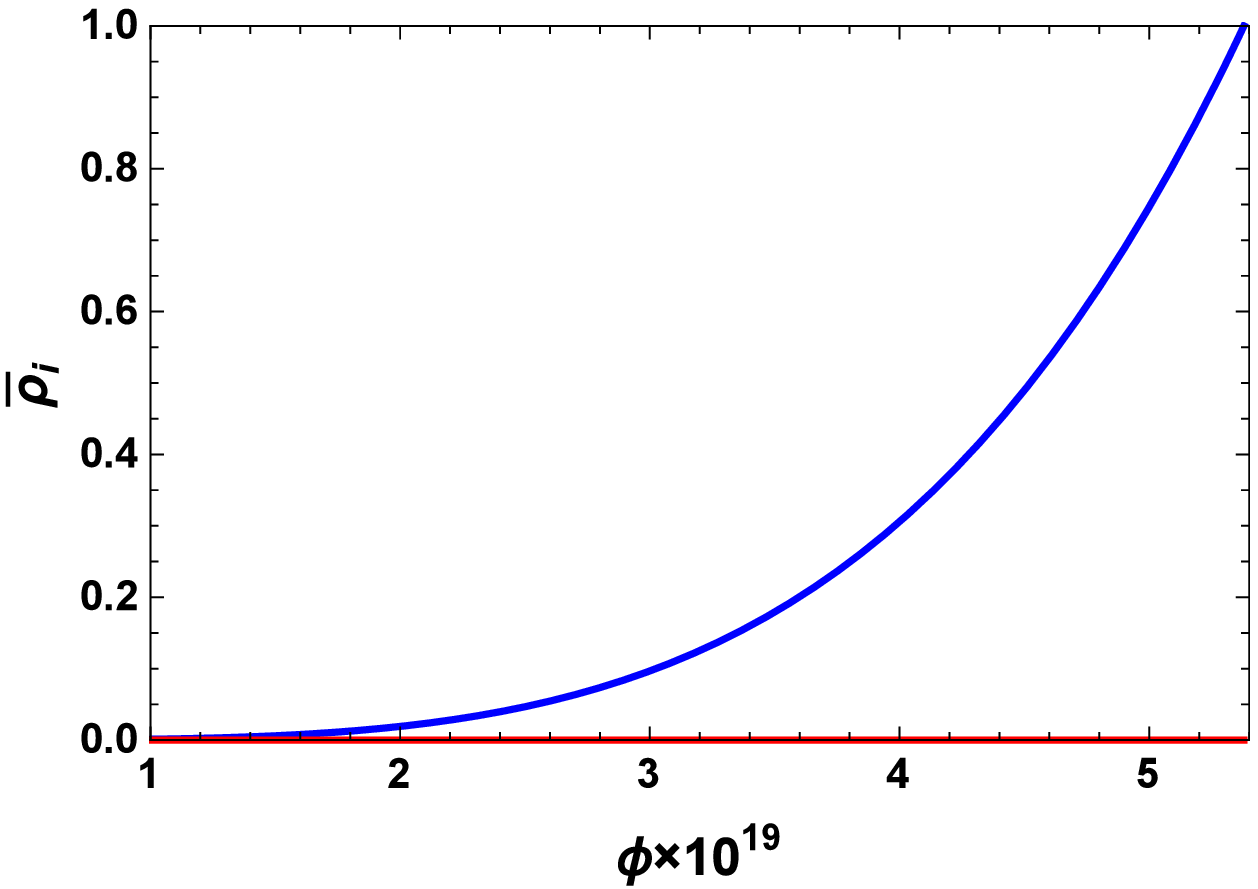}
  \caption{Energy densities of the scalar field (blue line) and radiation (red line) is depicted in term of the scalar field during the inflation times, where the constant parameters are taken as: $n=2$, $m=1$, $N=60$, and $H_0=1\times10^{-27}$. The parameter $\bar{\rho}_i$ is defined by $\bar{\rho}_i={\rho_i(\phi) \over \rho_0}$, in which $\rho_0$ is the initial value of the scalar field energy density.}\label{wenergy01}
\end{figure}

Computing the amplitude of scalar perturbation and the tensor-to-scalar ratio at horizon crossing, results in the following 
constraints for the free parameters
\begin{equation*}
  H_0 = \sqrt{\pi^2 M_p^2 r^\ast \mathcal{P}_s^\ast \over 2\left[ 2n^2 M_p^2 \Big( {2N \over n} + 1 \Big) \right]^n},
\end{equation*}
\begin{equation*}
  \Gamma_0 = {C \over H_0^5},
\end{equation*}
where the constant parameters $C$ is given by
\begin{equation*}
  C = \left[ 16 \alpha^{1/4} n^{3/2} M_p^3 \mathcal{P}_s^\ast \over 25\Big( 2n^2 M_p^2 \big( {2N \over n} +1\big) \Big)^{5n+m+6 \over 8} \right]^4.
\end{equation*}
Using these constraints, the potential scale and its shape could be estimated during inflation so that Fig.\ref{wpot01} clearly illustrates the scalar field potential. In contrast to the strong regime, the scalar field during the inflation is larger than the Planck mass. Also the potential is of order $10^{62}$ which is bigger than the potential in strong regime.
\begin{figure}
  \centering
  \includegraphics[width=7cm]{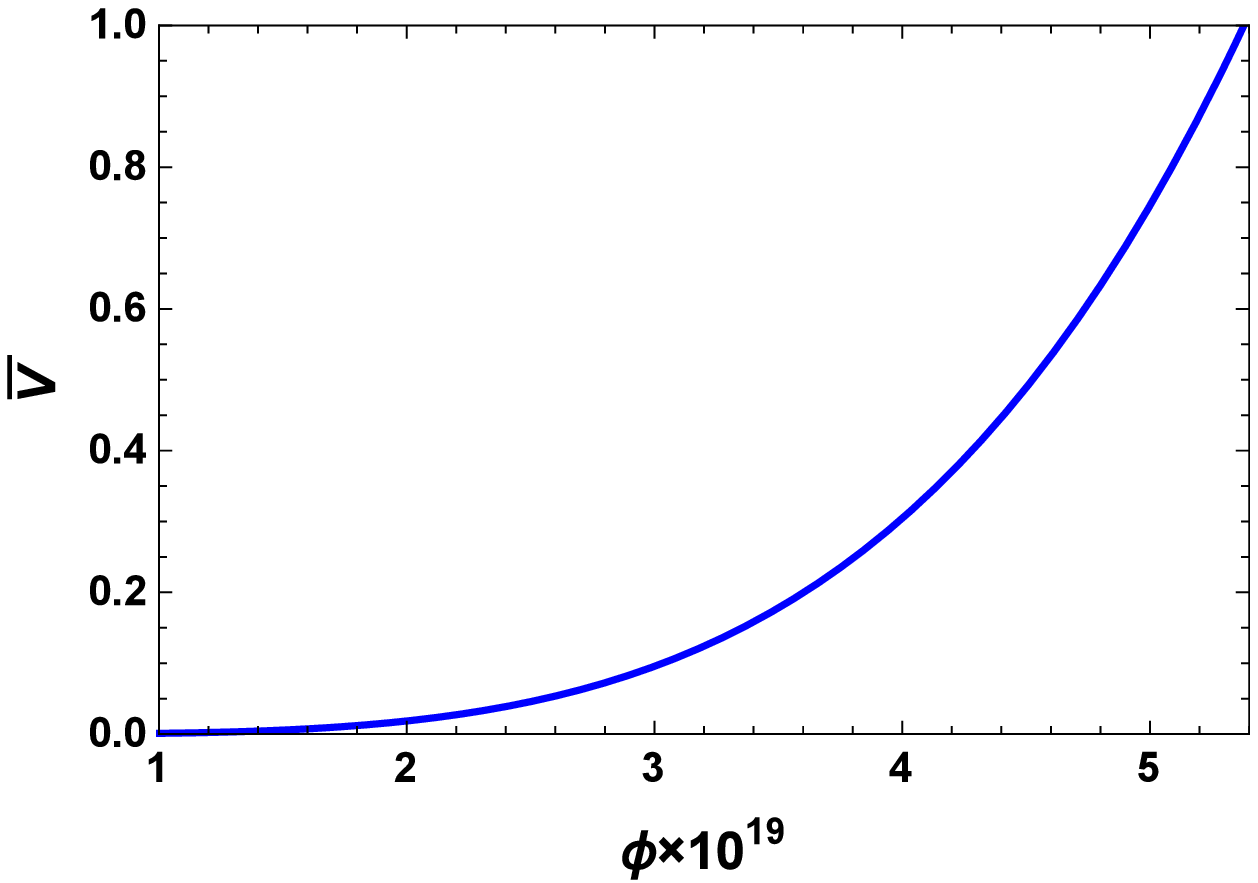}
  \caption{The potential versus the scalar field is plotted during the inflationary era, where the constant parameters are taken as: $n=2$, $m=1$, $N=60$, and $H_0=1\times10^{-27}$. The parameter $\bar{V}$ is defined by $\bar{V}={V(\phi) \over V_0}$, in which $V_0=1.48\times 10^{62}$ is the initial value of the potential.}\label{wpot01}
\end{figure}
On the other side, the constraints on the parameters could be used to compare the energy densities of the scalar field and radiation. Fig.\ref{wenergy01} portrays them versus the scalar field during the inflation, and it could be seen that the scalar field is strongly the dominant component of the Universe.

\subsection{$\Gamma := \Gamma(\phi,T)$}
\begin{figure}
  \centering
  \includegraphics[width=7cm]{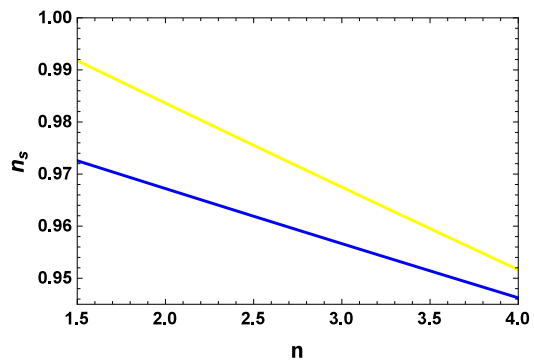}
  \caption{Scalar spectral index versus $n$ for $N=60$ and different values of $m$ as $m=1$ (blue color line) and $m=3$ (yellow color line).}\label{wns02}
\end{figure}

\begin{figure}
  \centering
  \includegraphics[width=7cm]{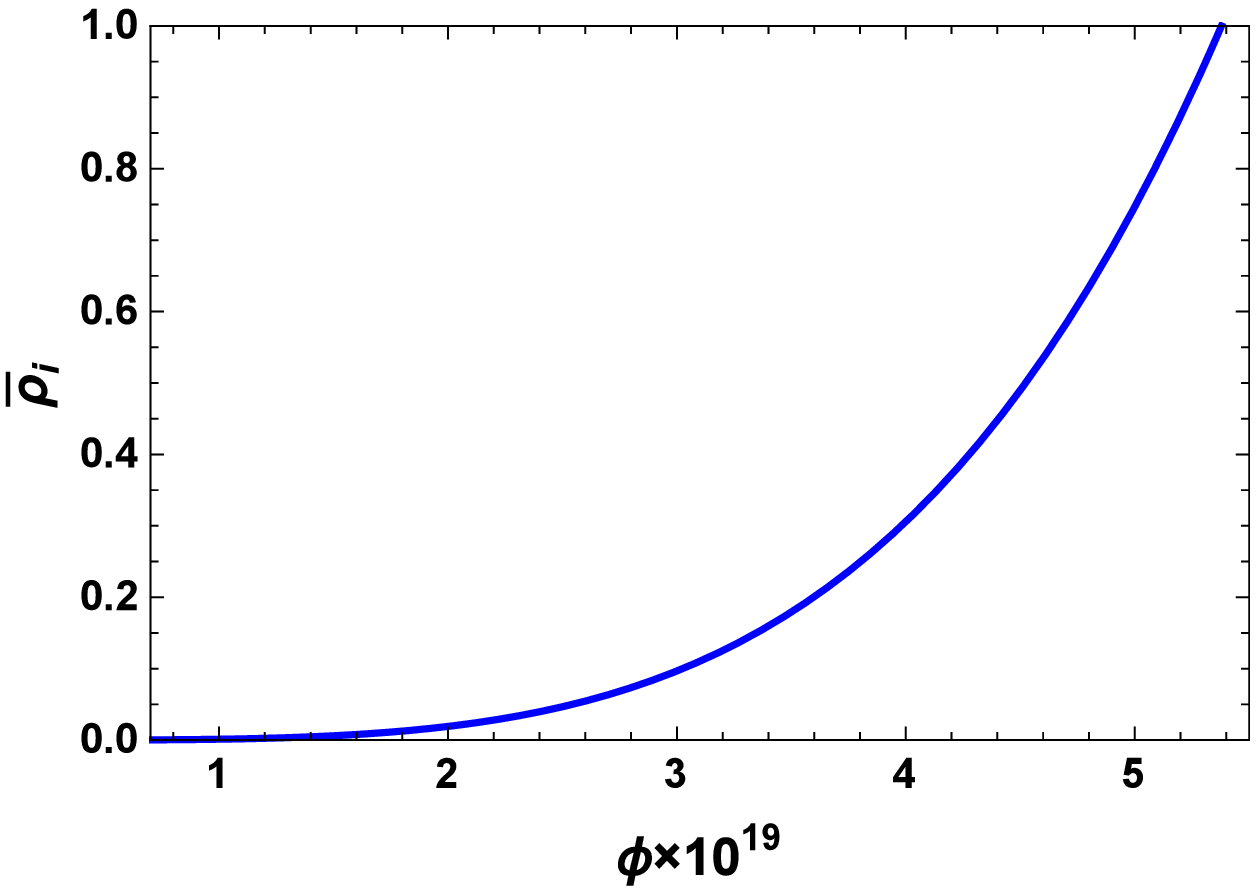}
  \caption{Energy densities of the scalar field (blue line) and radiation (red line) is depicted in term of the scalar field during the inflation times, where the constant parameters are taken as: $n=2$, $m=1.5$, $N=60$, and $H_0=1\times10^{-28}$. The parameter $\bar{\rho}_i$ is defined by $\bar{\rho}_i={\rho_i(\phi) \over \rho_0}$, in which $\rho_0$ is the initial value of the scalar field energy density. }\label{wenergy02}
\end{figure}

As the last case, the dissipative coefficient gets a more general form depending on both the scalar field and temperature, $\Gamma = \Gamma_0 {T^m \over \phi^{m-1}}$. Temperature could be read as a function of the scalar field from Eq.(\ref{rhor})
\begin{equation}\label{temp4}
  T = \left( {n^2 M_p^4 \over \alpha} \; H_0 \Gamma_0 \right)^{1 \over 4-m} \; \phi^{n-m-1 \over 4-m}.
\end{equation}
Therefore, the dissipative coefficient could be obtained in term of the scalar field
\begin{equation}\label{gamma4}
  \Gamma(\phi) = \left( {n^2 M_p^4 \over \alpha} H_0 \Gamma_0^{4 \over m} \right)^{m \over 4-m} \; \phi^{nm-6m+4 \over 4-m}.
\end{equation}
The slow-rolling parameters at the horizon exit have the same form as the previous case, and the only changes belongs to $\beta$ and $\delta$. Applying above dissipative coefficient, they are given by
\begin{eqnarray*}
\beta^\ast &=& {nm -6m+4 \over n(4-m)}\epsilon^\ast, \\
\delta^\ast &=& {(mn-5m)(nm -6m+4) \over \big( n(4-m) \big)^2} \epsilon^{\ast 2}.
\end{eqnarray*}
Consequently, the scalar spectral index and its running are derived as a function of $n$ and $m$. Depicting $n_s$ in term of $n$ for different values of $m$ in Fig.\ref{wns02}, clearly displays the appropriate values for the free parameters to get a suitable $n_s$. In oppose to the previous case, one could realized that for larger $m$, $n_s$ begins at bigger values and has larger slope.  \\
From above figure, it is believed that $(n,m)=(2,1.5)$ is a suitable choice that results in $n_s=0.9688$, and $\alpha_s=-5.10 \times 10^{-4}$, which shows that the model could estimates compatible result with Planck data. Illustrating the parameter $\alpha_s$ in term of $n_s$ for $N=55-65$ and standing the line in acceptable area determines the compatibility of the case with Planck data.
\begin{figure}
  \centering
  \includegraphics[width=7cm]{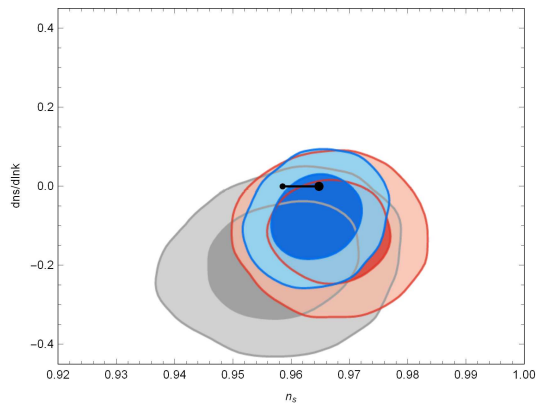}
  \caption{The running of scalar spectral index is shown versus $n_s$ for $n=2$ and $m=1.5$, and for $N=55-65$ in which the small point belongs to $N=55$ and the large point belongs to $N=65$.}\label{wasns01}
\end{figure}
The other two free parameters could be concluded by using data for the amplitude of scalar perturbation and the tensor-to-scalar ratio
\begin{equation*}
  H_0 = \sqrt{\pi^2 M_p^2 r^\ast \mathcal{P}_s^\ast \over 2\left[ 2n^2 M_p^2 \Big( {2N \over n} + 1 \Big) \right]^n},
\end{equation*}
\begin{equation*}
\Gamma_0 = {C \over H_0^{5-m}},
\end{equation*}
in which the constant $C$ is defined as
\begin{equation*}
C = \left[ {16 \alpha^{1/4} n^{3/2} M_p^3 \mathcal{P}_s^\ast \over 25\Big( {n^2 M_p^4 \over \alpha} \Big)^{m \over 4(4-m)} \Big( 2n^2M_p^2 \big( {2N \over n} + 1 \big)\Big)^{5n-nm-3m+7 \over 2(4-m)}} \right]^{4-m}.
\end{equation*}
Utilizing the appropriate and compatible values for the parameters, the potential could be figured out as in Fig.\ref{wpot01}. The scalar field is still larger than the Planck mass, however the potential remains below the Planck energy scale. Difference between the initial and final scalar field grows a little, and the potential behavior stays unchange where the scalar field slowly falls down from the top.
\begin{figure}
  \centering
  \includegraphics[width=7cm]{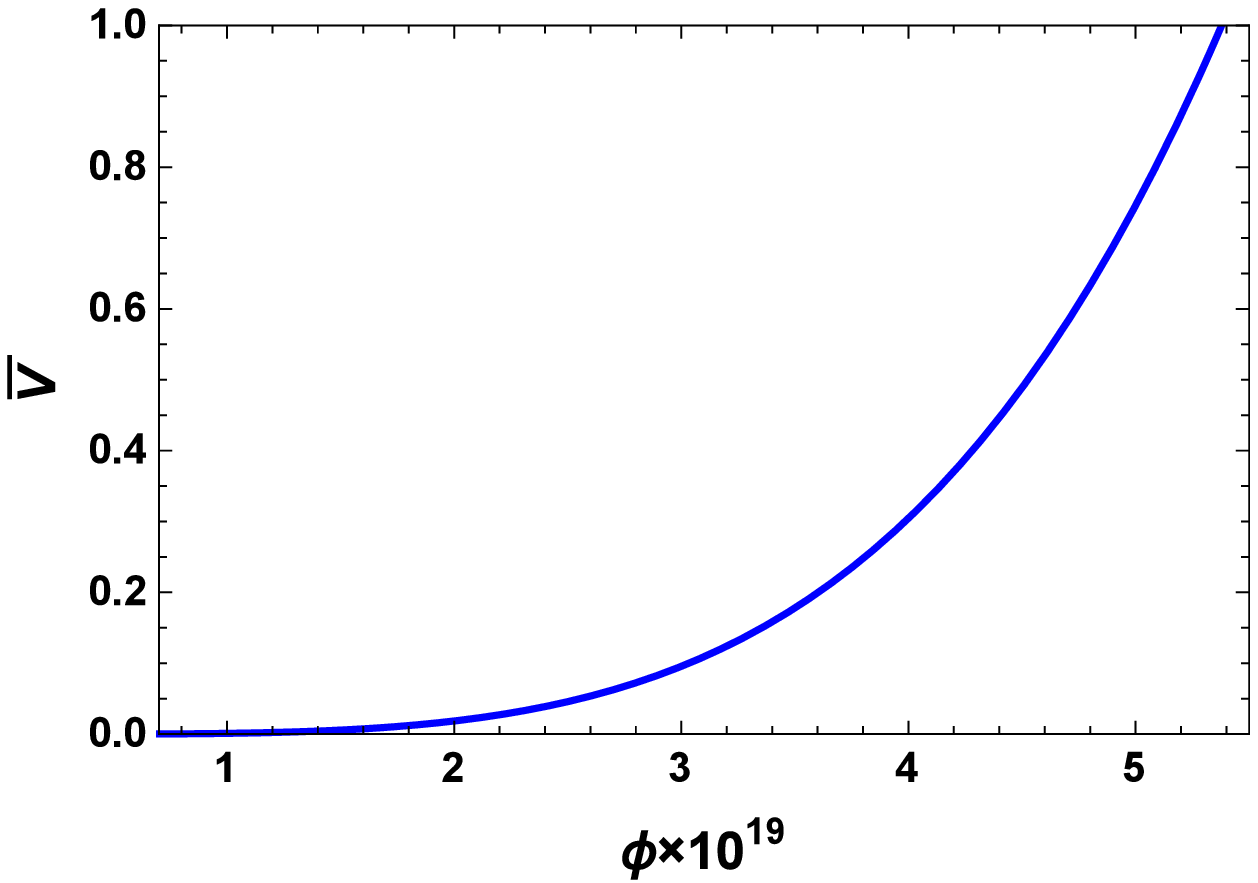}
  \caption{The potential versus the scalar field is plotted during the inflationary era, where the constant parameters are taken as: $n=2$, $m=1.5$, $N=60$, and $H_0=1\times10^{-28}$. The parameter $\bar{V}$ is defined by $\bar{V}={V(\phi) \over V_0}$, in which $V_0=1.48\times 10^{60}$ is the initial value of the potential.}\label{wpot01}
\end{figure}
Besides, the energy densities of the scalar field and radiation could be compared easily based on Fig.\ref{wenergy02} where one could find out the dominance of the scalar field during the inflation.


\section{Discussion and Conclusions}\label{conclusion}
Warm inflationary scenario as an alternative theory for the Universe evolution is the main interest of the presented work. The scalar field still plays the most important component that drives inflation, however by contrast to the cold inflation, is not the only component so that there are other fields interacting with the scalar field. The interaction is described by the dissipation coefficient appearing 
in the scalar field equation of motion. Then, temperature does not fall down dramatically such that at the end of inflation the 
Universe is able to smoothly enter to radiation era, and reheating phase could be avoided.   \\
Studying inflationary scenario is proceeded utilizing three common approached as: working with the potential, introducing scale 
factor, and proposing the Hubble parameter where the latter is known as Hamilton-Jacobi formalism. Here, some of these research works 
and their result will be discussed briefly, and it is clarified what choices they made for the essential quantities of the model. \\
In \cite{oliveira}, the authors first considered the perturbation in warm inflationary scenario, and then studied two typical examples 
as $V(\phi) = \lambda^4\phi^4 $ and $\Gamma = \Gamma_0 \phi^m$ for the first case, and for the second case they 
took $V(\phi) = m^2\phi^2 / 2$ and $\Gamma = \Gamma_0$. They proved that in both cases the scalar field is bigger than the 
Planck mass when the perturbation crosses the horizon, and at the end of inflation it goes below the Planck mass. Also they 
successfully constrained the constant coefficients of the potential and dissipation ratio. The scalar spectral index was 
derived as $n_s = 0.956$ for the first and $n_s=0.962$ for the second case. Besides, the temperature at the end of inflation 
was extracted as $T = 6.28 \times 10^{-8} \rm{m_p}$ and $ T = 0.52 \times 10^{-7} \rm{m_p}$ respectively for the first and 
second case.\\
In the recent work \cite{Visinella}, warm inflationary scenario was reconsidered by assuming a power-law function of scalar 
field for both the potential, $V(\phi) \propto \phi^p$, and dissipation coefficient, $\Gamma \propto \phi^{q/2}$. The work 
was restricted to strong dissipative regime. It was shown that the temperature at the end of inflation is about $10^{13} \rm{GeV}$ 
bigger than the Hubble parameter that was about $10^{10} \rm{GeV}$. The scalar field difference between initial and end of inflation 
was below the Planck mass. \\
Warm inflationary scenario in brane-world gravity with potential approach was considered in \cite{cid} by assuming a power-law 
function of scalar field for the potential, $V(\phi) \propto \phi^n$, and dissipation coefficient, $\Gamma \propto \phi^m$. It was 
found that for $n=4$ and $m=0$ the temperature at the end of inflation is of order $10^{-6} M_4$ and the scalar spectral index is 
very close to one as $n_s=0.997$. Also, for $n=6$ and $m=0$ the temperature was the same and scalar spectra index comes close to the 
recent observational data as $n_s = 0.966$. However, the point is that by this choice for $m$ the dissipation coefficient actually 
becomes a constant in both cases. Furthermore, the scalar field at the beginning of inflation is smaller than the four dimensional 
Planck mass as $\phi_\ast \propto 10^{-5} M_4$. \\
Warm intermediate inflation was considered in \cite{del-Campo}, where the authors took an exponential function of time for scale 
factor and the dissipation coefficient was assumed as $\Gamma = \Gamma_0 T^3 / \phi^2$. The work covered both strong and weak 
dissipative regimes, mostly concentrated on $r-n_s$ diagram. They could properly choose the free parameters of the model in a 
way to find result compatible with WMAP data for large number of e-folds as $N>140$. However, the obtained result seems not to 
be in good agreement with Planck data and in this regards the model needs to be reconsidered. \\
In the presented work, warm inflationary scenario was reconsidered utilizing a different approach named Hamilton-Jacobi formalism where instead of the scalar field potential, the Hubble parameter is given as a function of the scalar field. In order to go further, a power-law function of the scalar field is proposed for the Hubble parameter and the rest of the work is separated to two parts as strong and weak dissipative regimes, where two popular forms of the dissipation coefficient as $\Gamma(\phi)=\Gamma_0\phi^m$ and the general function $\Gamma(\phi,T)=\Gamma_0 {T^m \over \phi^{m-1}}$ were considered in great detail for each regime.  \\
In strong dissipative regime, a suitable choice of the model parameters $(n,m)$ resulted in an acceptable values for the scalar spectral index and its running for both choices of $\Gamma$ so that they stood in perfect agreement with the Planck data. However one still could get more insight by illustrating the running of scalar spectral index versus the scalar spectral index and fit the figure on Planck data. By doing so, the result was compatible with the data that in turn displayed positive side of the model. On the other side, using the amplitude of scalar perturbation and the tensor-to-scalar ratio relations and applying the corresponding data, the other two free parameters could be restricted too. A way of testing the constraint is plotting $r-n_s$ graph for all cases, as they have been determined in Fig.\ref{rns}. It is clearly seen that the model prediction about the perturbation parameters stands in $68\%$ CL area which in turn shows a perfect agreement with Planck data. In another words, the constraints on the free parameters have been imposed properly. \\
Since there is only an upper bound for $r^\ast < 0.11$ it sounds that the parameter $H_0$ still has a freedom, however it should be noted that the parameter should be chosen in a way to result a potential below the Planck energy scale and a temperature higher than the Hubble parameter (to satisfy the warm inflationary condition). Satisfying these conditions is very important that sometimes is forgotten. Constraining the parameters shows that the scalar field during the inflationary times could go below the Planck mass, in which for the second case the scalar field is smaller than the first scale. Besides, the potential remains below the Planck energy scale as well. Depicting the potential illustrated that during the inflation it has the similar shape as chaotic potential so that the initial scalar field stands on top of its potential and slowly rolls down to the minimum. On the other hand, warm inflation condition $(T>H)$ has been examined by plotting the parameter versus the scalar field during the inflation time in Fig.\ref{temp} that displays that the condition is satisfied successfully. \\
Additionally, the dominance of the scalar field energy density over the radiation was examined, and it was exhibited that the free parameters were selected properly.  \\
Almost the same result was concluded for the weak dissipative regime: the perturbation parameters were estimated standing in the data range. Although the potential was derived below the Planck energy scale, the scalar field during the inflation era was bigger than the Planck mass.\\

\newpage

\begin{widetext}

\begin{figure}[ht]
  \centering
  \subfigure[strong-first case]{\includegraphics[width=5.5cm]{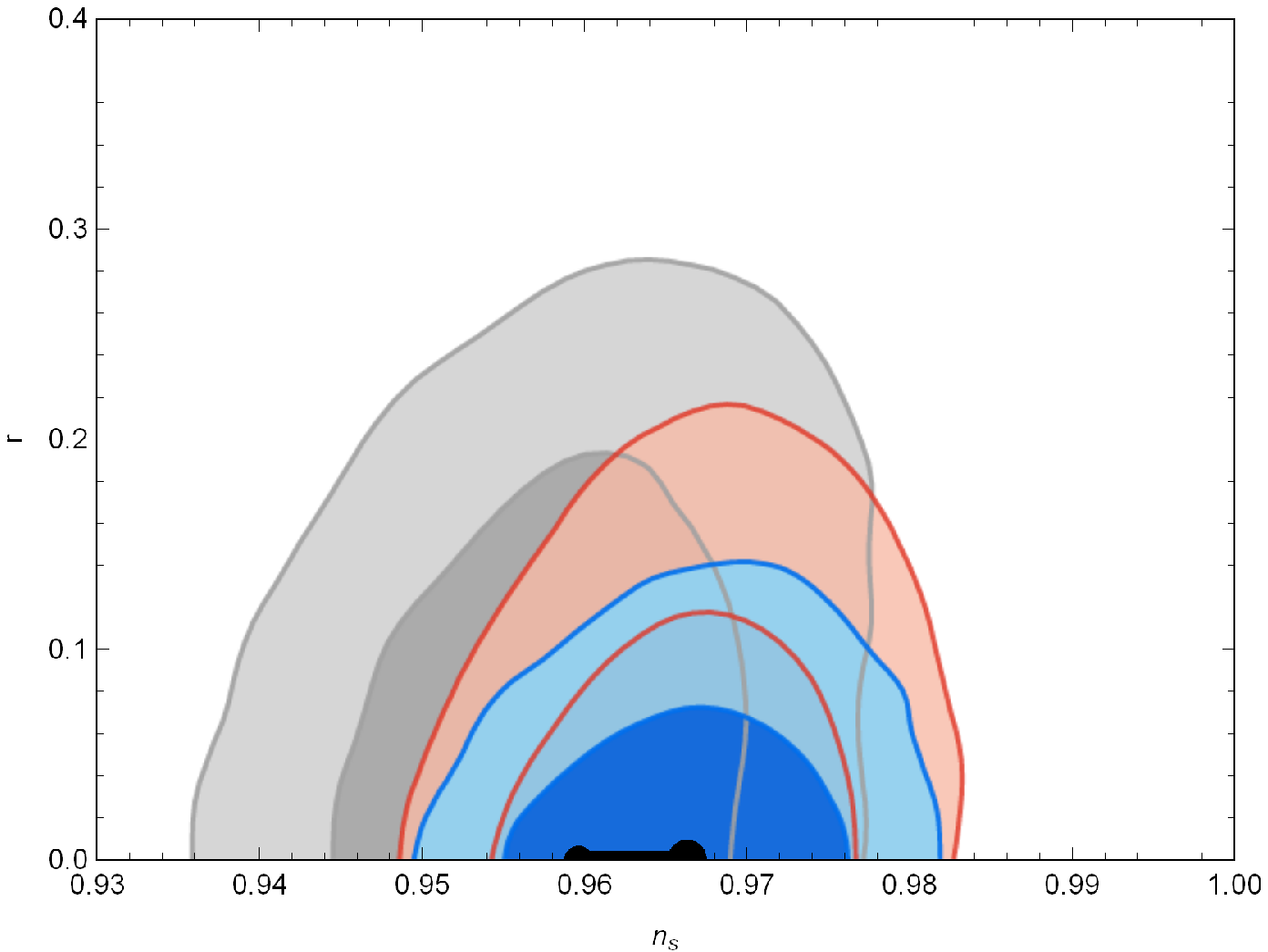}}\hspace{1cm}
  \subfigure[strong-first case]{\includegraphics[width=5.5cm]{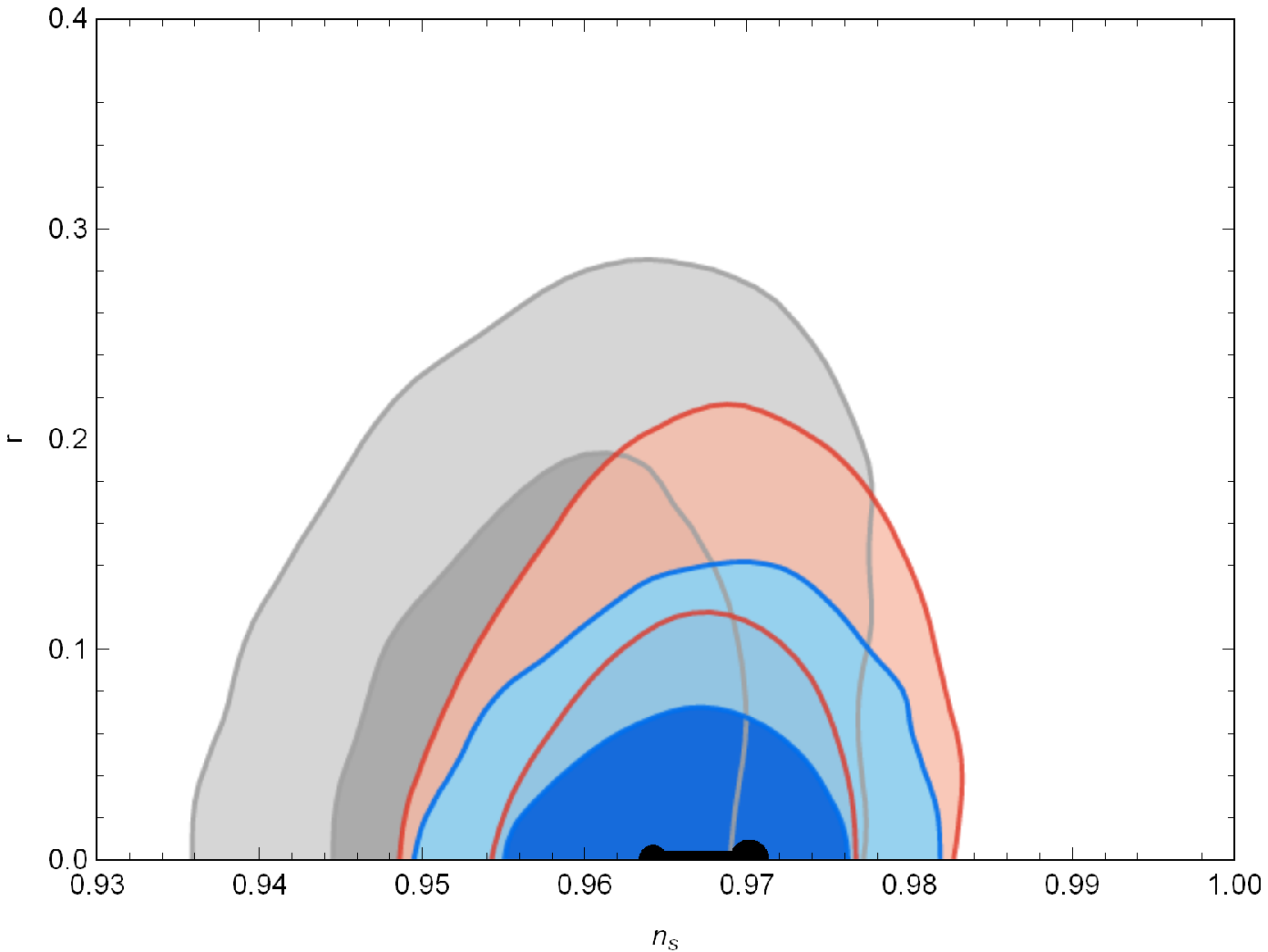}}\hspace{1cm}
  \subfigure[weak-first case]{ \includegraphics[width=5.5cm]{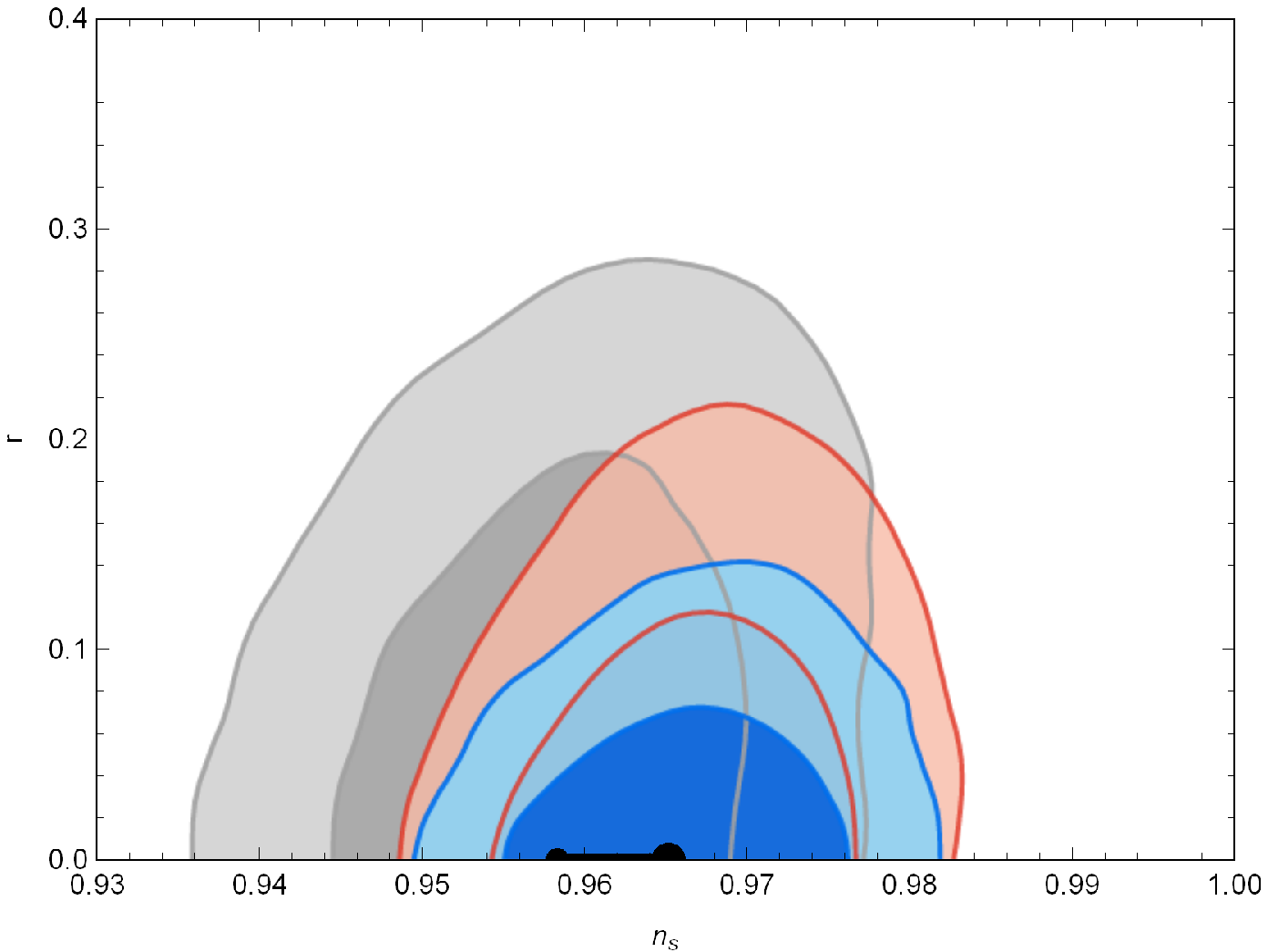}}\hspace{1cm}
  \subfigure[weak-first case]{ \includegraphics[width=5.5cm]{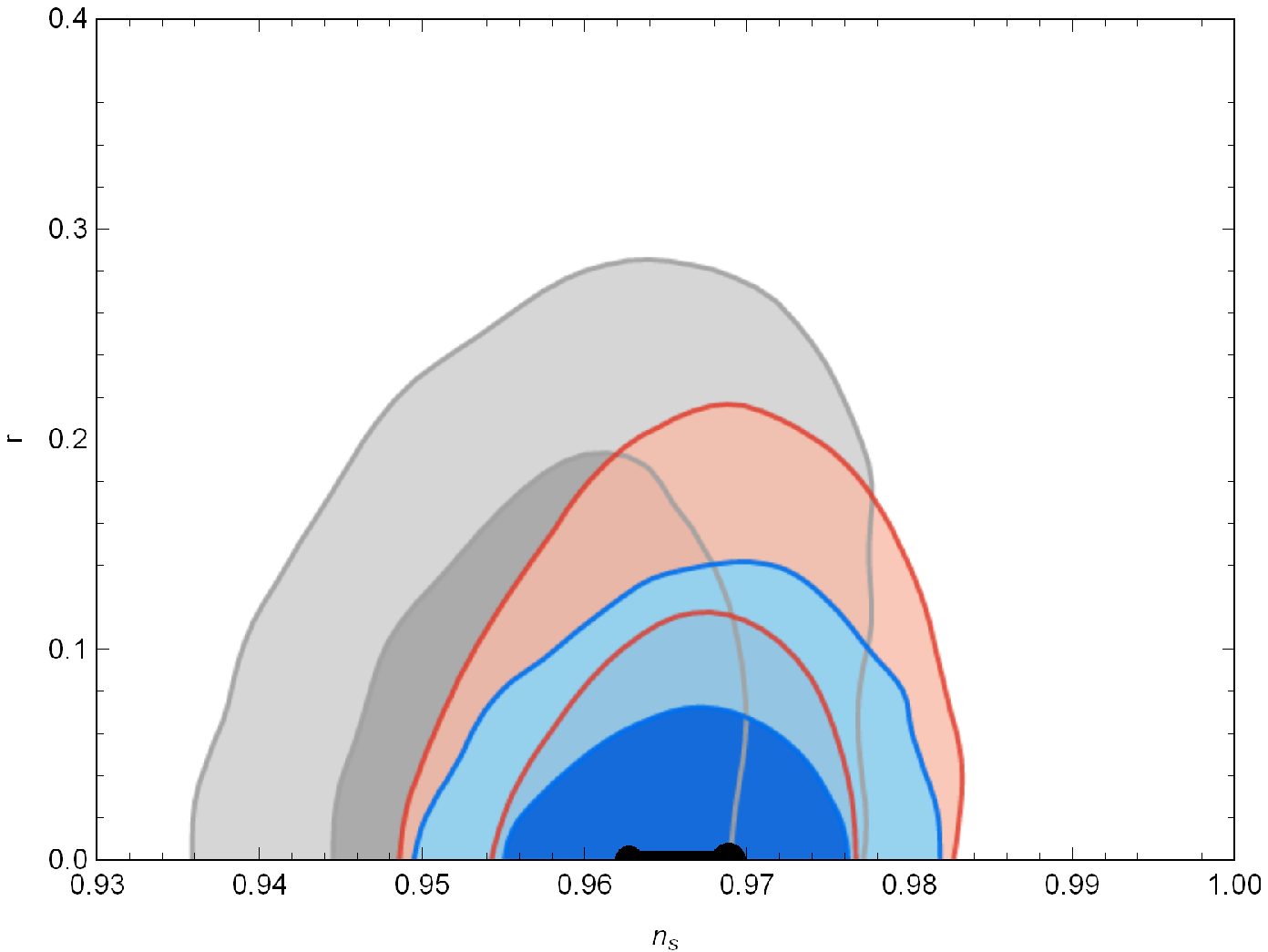}}
  \caption{$r-n_s$ diagram for selected free parameters have been plotted for every cases in which the small point is related to $N=50$ and the large point corresponds to $N=60$. }\label{rns}
\end{figure}


\begin{figure}[h]
  \centering
  \subfigure[strong-first case]{\includegraphics[width=5.5cm]{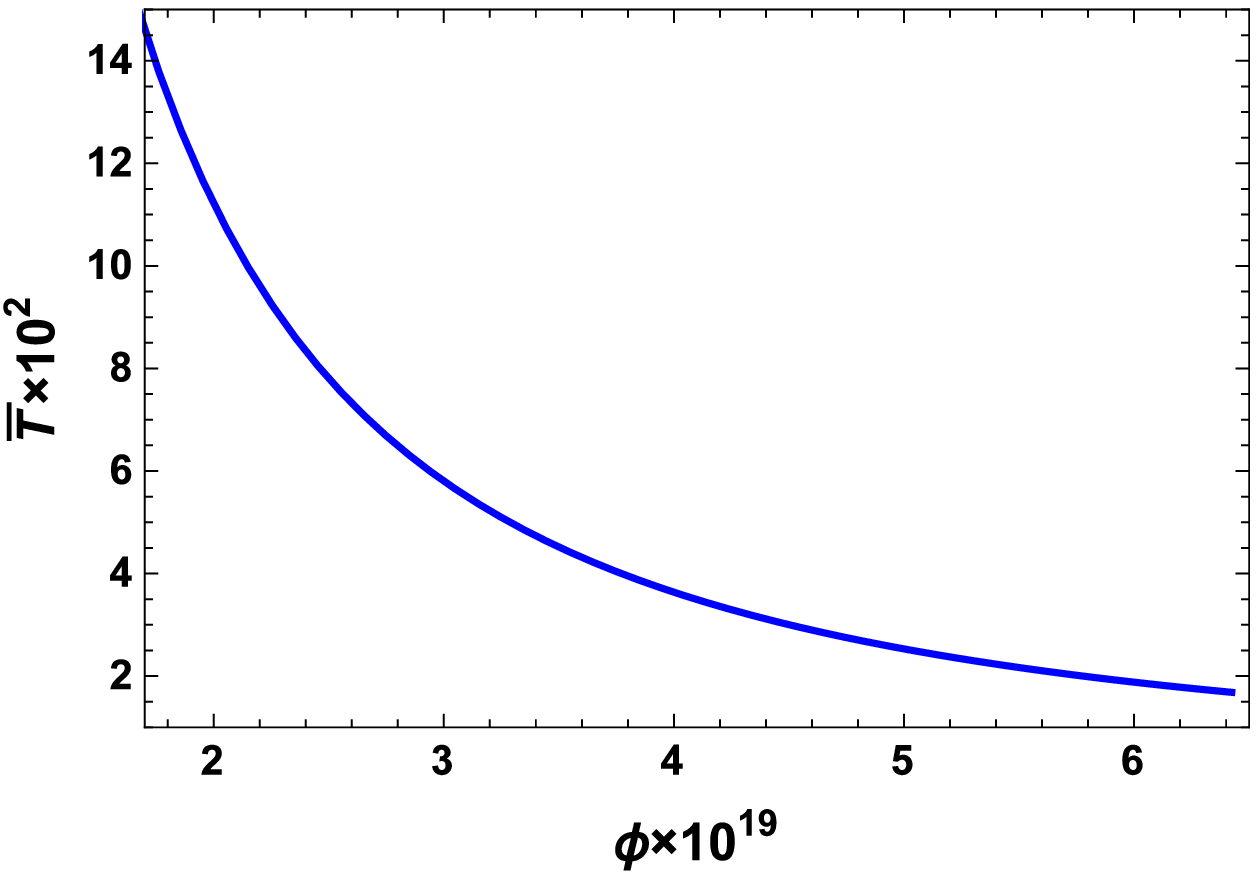}} \hspace{1cm}
  \subfigure[strong-first case]{\includegraphics[width=5.5cm]{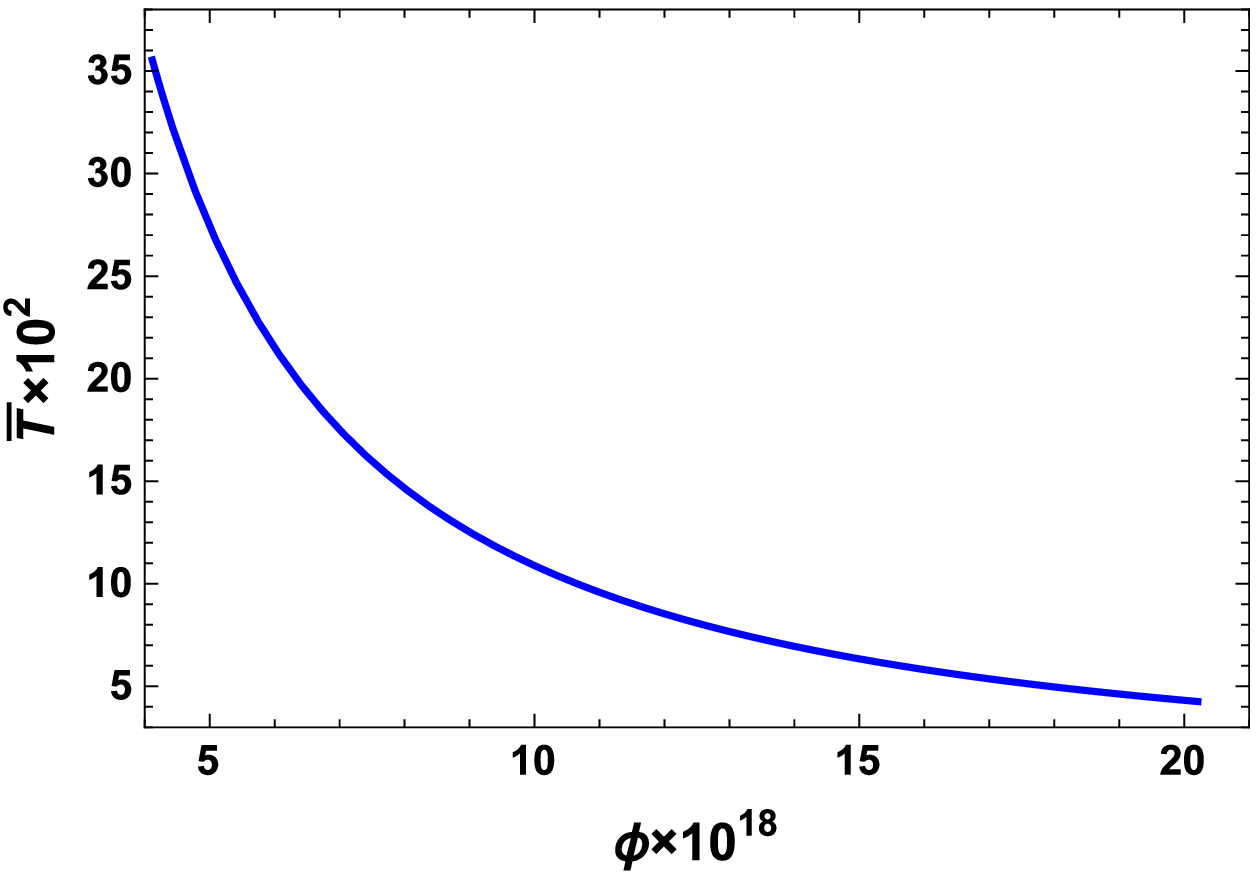}}\hspace{1cm}
  \subfigure[weak-first case]{ \includegraphics[width=5.5cm]{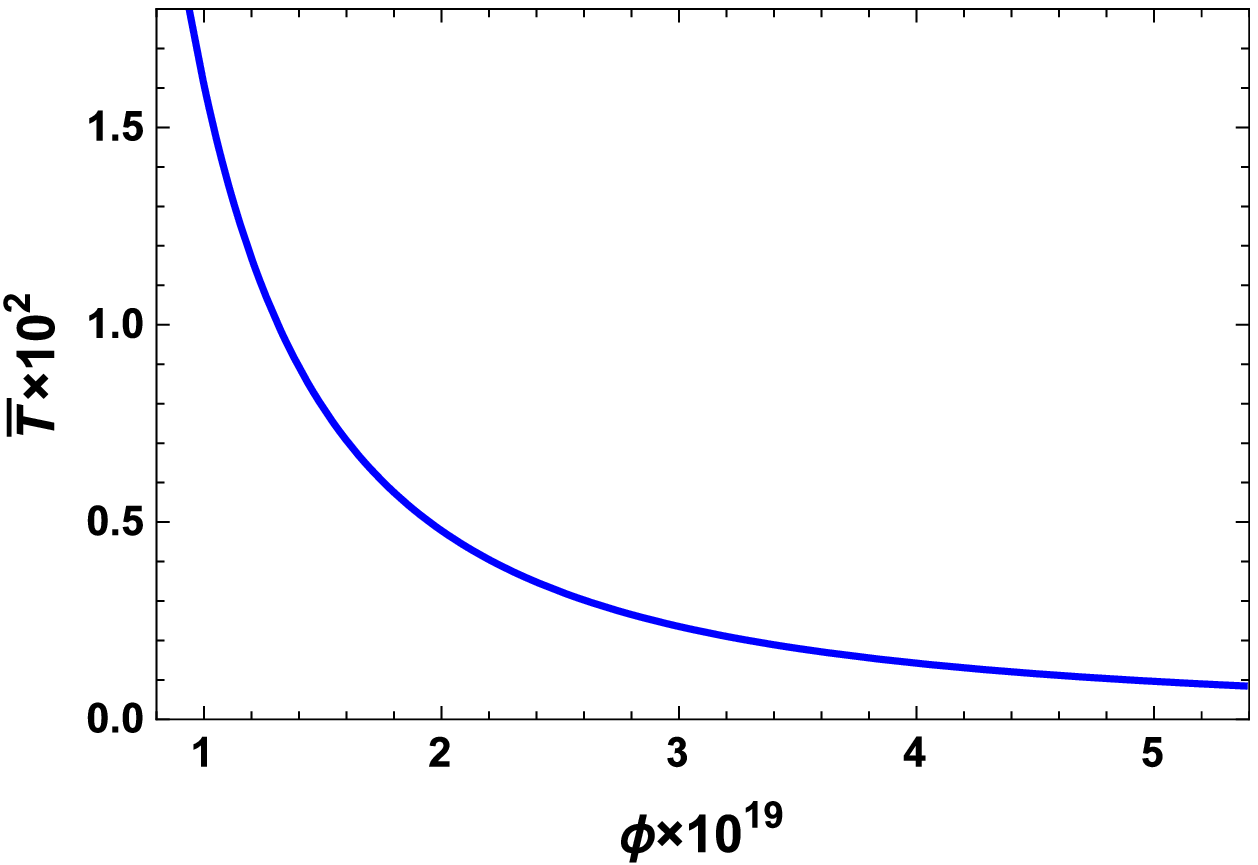}}\hspace{1cm}
  \subfigure[weak-first case]{ \includegraphics[width=5.5cm]{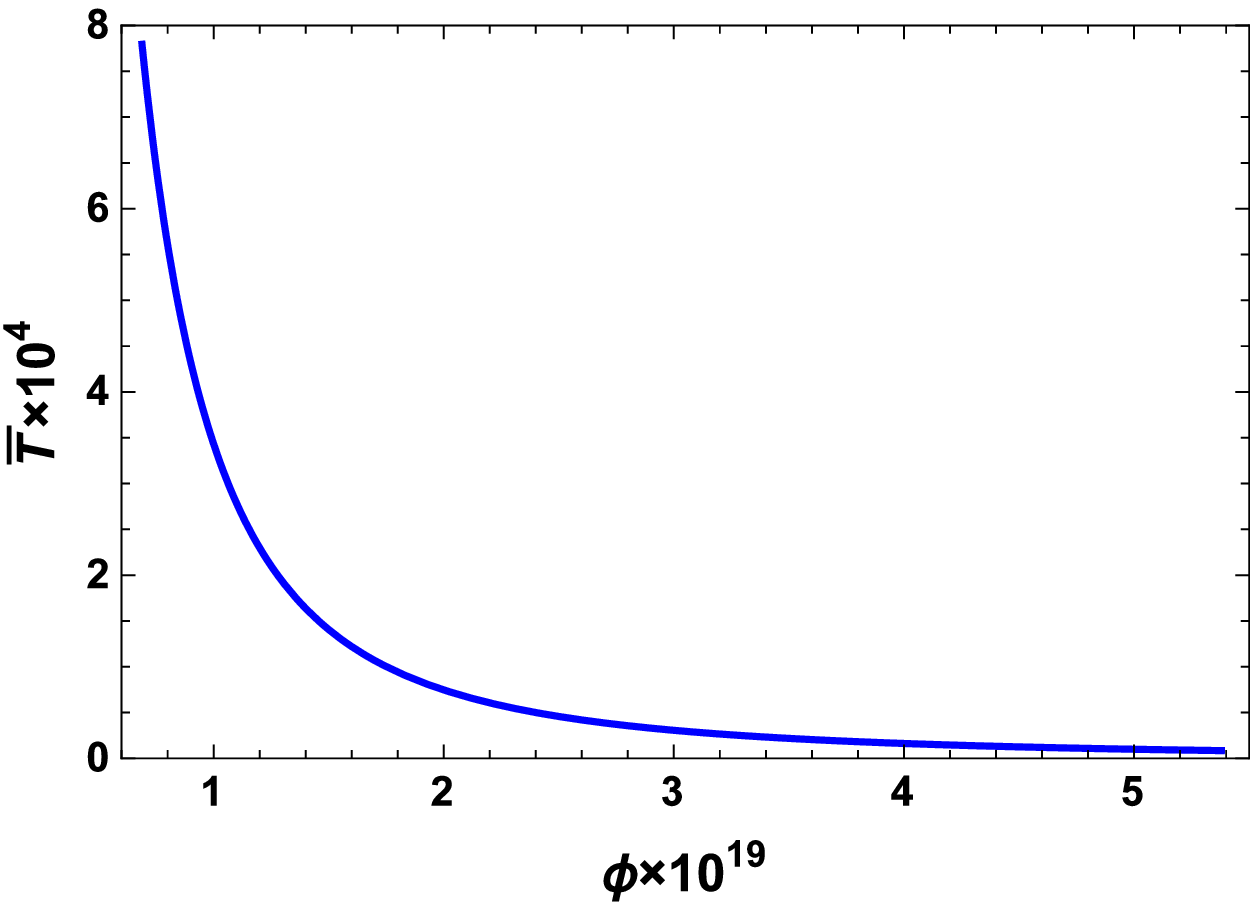}}
  \caption{Temperature versus the scalar field during the inflationary times has been plotted for all cases, for the corresponding chosen free parameters. The parameter $\bar{T}$ is defined as $\bar{T}=T/H$}\label{temp}
\end{figure}

\end{widetext}










\end{document}